\documentclass[11pt,letterpaper]{article}
\pdfoutput=1
\usepackage{jcappub}
\usepackage{color}
\usepackage{graphicx}

\usepackage{verbatim}
\usepackage{amsmath}
\usepackage{amssymb}
\usepackage{subfig}
\usepackage{url}
\usepackage{bbold}

\usepackage{multirow}
\usepackage{threeparttable}
\usepackage{paralist}

\newcommand{\GeV}{\text{GeV}}

\newcommand{\SU}{\text{SU}}

\newcommand{\U}{\text{U}}

\newcommand{\vev}[1]{\langle #1 \rangle}

\DeclareRobustCommand{\Sec}[1]{Sec.~\ref{#1}}
\DeclareRobustCommand{\Secs}[2]{Secs.~\ref{#1} and \ref{#2}}

\DeclareRobustCommand{\Tab}[1]{Table~\ref{#1}}

\DeclareRobustCommand{\Fig}[1]{Fig.~\ref{#1}}

\DeclareRobustCommand{\Eq}[1]{Eq.~(\ref{#1})}

\DeclareRobustCommand{\Ref}[1]{Ref.~\cite{#1}}
\DeclareRobustCommand{\Refs}[1]{Refs.~\cite{#1}}

\newcommand{\be}{\begin{equation}}
\newcommand{\ee}{\end{equation}}

\usepackage{xspace}

\newcommand{\Fermi}{\textit{Fermi}\xspace}
\newcommand{\alphaEM}{\alpha_{\rm EM}}
\newcommand{\alphaD}{\alpha_d}
\newcommand{\Gdark}{G_d}
\newcommand{\Hdark}{H_d}

\def\Tr{\mathop{\rm Tr}}

\begin{document}

\title{Multiple Gamma Lines from Semi-Annihilation}
 
\author[a,b,c]{Francesco D'Eramo,}
\author[c]{Matthew McCullough,}
\author[c]{and Jesse Thaler}

\affiliation[a]{Department of Physics, University of California, Berkeley, CA 94720, USA}
\affiliation[b]{Theoretical Physics Group, Lawrence Berkeley National Laboratory, Berkeley, CA 94720, USA}
\affiliation[c]{Center for Theoretical Physics, Massachusetts Institute of Technology, Cambridge, MA 02139, USA}

\emailAdd{fraderamo@berkeley.edu}
\emailAdd{mccull@mit.edu}
\emailAdd{jthaler@mit.edu}

%\date{\today}

\abstract{Hints in the \Fermi data for a $130~\GeV$ gamma line from the galactic center have ignited interest in potential gamma line signatures of dark matter.  Explanations of this line based on dark matter annihilation face a parametric tension since they often rely on large enhancements of loop-suppressed cross sections.  In this paper, we pursue an alternative possibility that dark matter gamma lines could arise from ``semi-annihilation'' among multiple dark sector states.  The semi-annihilation reaction $\psi_i \psi_j \to \psi_k \gamma$ with a single final state photon is typically enhanced relative to ordinary annihilation $\psi_i \overline{\psi}_i \to \gamma \gamma$ into photon pairs.  Semi-annihilation allows for a wide range of dark matter masses compared to the fixed mass value required by annihilation, opening the possibility to explain potential dark matter signatures at higher energies.  The most striking prediction of semi-annihilation is the presence of multiple gamma lines, with as many as order $N^3$ lines possible for $N$ dark sector states, allowing for dark sector spectroscopy.  A smoking gun signature arises in the simplest case of degenerate dark matter, where a strong semi-annihilation line at $130~\GeV$ would be accompanied by a weaker annihilation line at $173~\GeV$.  As a proof of principle, we construct two explicit models of dark matter semi-annihilation, one based on non-Abelian vector dark matter and the other based on retrofitting Rayleigh dark matter.}

%\arxivnumber{XXXX.XXXX}

\preprint{MIT-CTP {4408} \hspace{0.2cm} UCB-PTH-12/18}

\maketitle

\section{Introduction}
\label{sec:introduction}

While the gravitational evidence for dark matter (DM) is overwhelming \cite{Jungman:1995df,Bergstrom:2000pn,Bertone:2004pz,Ellis:2010kf,Feng:2010gw}, the precise nature of DM is still unknown.  An attractive and well-studied possibility is for DM to be composed of long-lived neutral particles (WIMPs)  \cite{Lee:1977ua,Scherrer:1985zt,Kolb:1985nn,Srednicki:1988ce,Gondolo:1990dk}.  In such scenarios, one typically expects feeble non-gravitational interactions between DM and standard model (SM) particles, especially if the relic density of DM is determined by thermal freeze out.  Such interactions allow DM in the galactic halo to annihilate or decay to SM states, motivating a variety of DM indirect detection experiments.

Recent hints of a 130 GeV gamma line in the \Fermi public data \cite{Bringmann:2012vr,Weniger:2012tx,Su:2012ft,Tempel:2012ey,Su:2012zg} have highlighted the importance of galactic photons as a probe of DM.  Since standard astrophysical processes do not yield monochromatic photons at these high energies,\footnote{See \Ref{Aharonian:2012cs} for a discussion of potential astrophysical sources of gamma lines.} the presence of line emission from the galactic center is an unusually robust probe of DM \cite{Bergstrom:1988fp,Rudaz:1989ij,Bergstrom:1997fh,Ullio:1997ke,Bern:1997ng,Bergstrom:2004nr,Bringmann:2012ez}.  The morphology of the gamma line feature is consistent with DM annihilation \cite{Rajaraman:2012db,Hektor:2012kc,Hooper:2012qc,Finkbeiner:2012ez,Rao:2012fh}, in which case the measured photon energy is equal to the DM mass.  Assuming a standard DM halo profile, the annihilation cross section is consistent with $\langle \sigma v\rangle_{\chi\chi \rightarrow \gamma \gamma} \simeq 1.3\times10^{-27}\, {\rm cm}^3\, {\rm s}^{-1}$ \cite{Weniger:2012tx}, corresponding to around one-thirtieth of the expected thermal freeze out cross section.  This tentative gamma line feature has motivated a number of new DM scenarios, based on both DM annihilations \cite{Dudas:2012pb,Cline:2012nw,Choi:2012ap,Lee:2012bq,Buckley:2012ws,Weiner:2012cb,Chu:2012qy,Kang:2012bq,Tulin:2012uq,Cline:2012bz,Bai:2012qy,Bergstrom:2012bd,Wang:2012ts,Weiner:2012gm,Lee:2012wz,Fan:2012gr} and DM decays \cite{Park:2012xq,Kyae:2012vi}, as well as constraints on such DM models from accompanying continuum photon emission \cite{Buchmuller:2012rc,Cohen:2012me,Cholis:2012fb,Blanchet:2012vq,Huang:2012yf}.  While alternative interpretations of the \Fermi line have been put forward \cite{Profumo:2012tr,Boyarsky:2012ca,Mirabal:2012za,Hektor:2012jc,Whiteson:2012hr,Hektor:2012ev}, the DM hypothesis is still very appealing, and the DM community eagerly awaits an independent analysis by the \Fermi collaboration.

In this paper, we present an alternative interpretation of the 130 GeV gamma line as arising from ``semi-annihilation'' of DM  \cite{D'Eramo:2010ep} (see also \Refs{Hambye:2008bq,Hambye:2009fg,Arina:2009uq}).  Semi-annihilation appears in models with multiple DM species, and  creates a gamma line through the process
\be
\label{eq:semigamma}
\psi_i \psi_j \rightarrow \psi_k \gamma,
\ee
where $\psi_i$ corresponds to the $i$-th DM particle.  Here, the monochromatic photon energy is given by the combination
\be
E^{ij\to k}_\gamma = \frac{(m_i + m_j + m_k) (m_i + m_j - m_k)}{2 (m_i + m_j)},
\ee
where $m_i$ is the mass of $\psi_i$.  In particular, if all DM species have the same mass, then a 130 GeV line corresponds to 173 GeV DM.  More generally, semi-annihilation predicts a corresponding line for each $ij \to k$ trio.

Semi-annihilation possesses a number of features which make it interesting to study, even if the present \Fermi hints ultimately have a non-DM explanation.  
\begin{itemize}
\item \textbf{Parametrically larger cross section}.  For neutral DM, the annihilation process $\psi \overline{\psi} \rightarrow \gamma \gamma$ is generated by loop diagrams with an $\alpha_{\rm EM}^2$ suppression factor,\footnote{See however \Ref{Fan:2012gr} for a tree-level explanation.} making it generically too small to account for the large $\langle \sigma v\rangle_{\chi\chi \rightarrow \gamma \gamma} \simeq 1.3 \times 10^{-27}\, {\rm cm}^3\, {\rm s}^{-1}$ cross section required to explain the 130 GeV feature.  While semi-annihilation diagrams are also loop suppressed, $\psi_i \psi_j \rightarrow \psi_k \gamma$ is only proportional to a single factor of $\alpha_{\rm EM}$, yielding a modest enhancement relative to the annihilation case.
\item \textbf{Wide range of dark matter masses}.  In contrast to annihilation, gamma lines from semi-annihilation do not uniquely fix the DM mass.  Therefore a signal at $130~\GeV$ could arise from a wide range of DM masses.  While not the focus of this work, this could allow a common DM explanation for both the $130~\GeV$ gamma ray line and electron/positron excesses observed at higher energies \cite{Adriani:2008zr,Aharonian:2008aa,Abdo:2009zk}.  Of course, one must account for the fact that at fixed DM density $\rho_{\rm DM} = m_{\rm DM} n_{\rm DM}$, the indirect detection rate scales as $n_{\rm DM}^2 \propto 1/m_{\rm DM}^2$.
\item \textbf{Accompanying annihilation signal}.  Generic models of semi-annihilation do predict an annihilation signal $\psi_i \overline{\psi}_i \rightarrow \gamma \gamma$, albeit of weaker strength.  In the case of degenerate DM states, the 130 GeV semi-annihilation line would be accompanied by a weaker 173 GeV annihilation line, and detecting such a line would be a smoking gun for the semi-annihilation process.  More generally, the $E^{ij\to k}_\gamma$ line should have $E^i_\gamma$, $E^j_\gamma$, and $E^k_\gamma$ companions, offering the possibility of DM spectroscopy.  
\item \textbf{Generic absence of a 112 GeV feature}.   The $\psi \overline{\psi} \rightarrow \gamma \gamma$ annihilation diagram (with $E_\gamma = 130~\GeV$) usually has a companion diagram $\psi \overline{\psi} \rightarrow \gamma Z$ (with $E_\gamma = 112~\GeV$), and there are tentative indications of this second line in the data \cite{Su:2012ft,Su:2012zg,Rajaraman:2012db}.  Indeed, if the $112~\GeV$ feature were not to persist, this would disfavor the annihilation interpretation since most annihilation scenarios yield two lines (see \Ref{Rajaraman:2012db}).  In the case of semi-annihilation, the companion diagram to $\psi_i \psi_j \rightarrow \psi_k \gamma$ is $\psi_i \psi_j \rightarrow \psi_k Z$ which has no photon in the final state, and therefore no corresponding $112~\GeV$ feature.  Of course, there may be an $ij \to k$ trio that ``accidentally'' yields a $112~\GeV$ line, but this is not a robust prediction of the semi-annihilation framework.  
\end{itemize}
Given these unique features, it is worthwhile to consider various models and implications of semi-annihilation.

 \begin{figure}
\begin{center}
\includegraphics[scale=0.55]{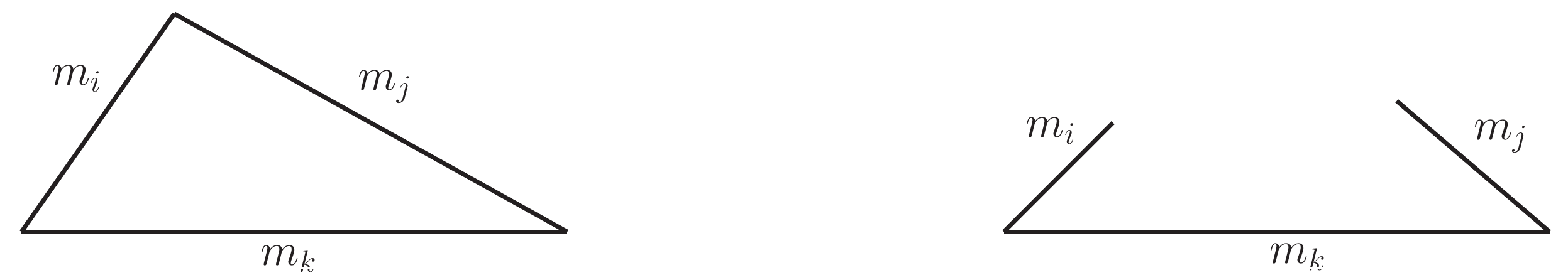} 
\caption{Two different spectra for the DM species $\psi_i$, adapted from \Ref{D'Eramo:2010ep}. In the case on the left, the three particles are mutually stable since their masses satisfy the triangle inequality.  In the case on the right, the decay $\psi_k \rightarrow \psi_i  \psi_j \phi^*$ is kinematically allowed.}
\label{fig:triangle}
\end{center}
\end{figure}

The remainder of this paper is organized as follows.  In \Sec{sec:linesfromsemi}, we review the semi-annihilation framework and explain the generic physics that leads to gamma lines.  We present a new semi-annihilation model based on non-Abelian vector DM in \Sec{sec:model}, and show how to retrofit existing annihilation models to have enhanced semi-annihilation in \Sec{sec:retrofit}.  We conclude in \Sec{sec:conclude}.

\section{Gamma Lines and Semi-Annihilation}
\label{sec:linesfromsemi}

The most well-studied models of WIMP DM involve a single relic particle stabilized by a $Z_2$ symmetry, as for example in supersymmetric models with $R$-parity.  However, any discrete or continuos global symmetry can be used for DM stabilization, opening the possibility for a far richer dark sector with multiple stable particles.  Here, we focus on stabilization symmetries that allow for the semi-annihilation reaction, and consider the implications for indirect DM detection.  We also briefly mention collider and direct detection expectations.

\subsection{Review of Semi-Annihilation}

For a generic dark sector with multiple stable species $\psi_i$, the semi-annihilation reaction reads~\cite{D'Eramo:2010ep}
\be
\psi_i \psi_j \rightarrow \psi_k \phi,
\label{eq:semi}
\ee
where $\psi_i$ are stable DM particles, and $\phi$ is either a SM field or a dark sector field which eventually decays to the SM.  This reaction can take place if the DM is stabilized by a symmetry larger than just $Z_2$.\footnote{The simplest example is a $Z_3$ symmetry, where DM is composed of a single stable component $\psi$ with a semi-annihilation reaction $\psi \psi \rightarrow \overline{\psi} \phi$.}  Such reactions are ubiquitous in multi-component models where the different species $\psi_i$ are stabilized by ``baryon'' and/or ``flavor'' symmetries, as in QCD-like theories.  In semi-annihilation, the total DM number changes by only one unit, in contrast to ordinary annihilation where the total DM number changes by two units.

The reaction in \Eq{eq:semi} could potentially make one of the DM species unstable, since the decay process $\psi_k \rightarrow \psi_i  \psi_j \phi^*$ can be obtained by crossing symmetry.  However, such decays may be kinematically forbidden.  For example, in the absence of the weak interaction, the proton $p$, neutron $n$, and charged pion $\pi^\pm$ are mutually stable, and the process $p \pi^- \to n \pi^0$ with $\pi^0 \to \gamma \gamma$ would be an example of semi-annihilation.  More generally, as long as the DM spectrum satisfies the triangle inequality $m_k <  m_i +m_j$, as well as its crossed versions, the semi-annihilation reaction does not destabilize DM.   This condition on the DM spectrum is depicted in \Fig{fig:triangle}.

When semi-annihilation is present, DM dynamics in the early universe are far more varied than in the standard $Z_2$ case~\cite{Hambye:2008bq,Hambye:2009fg,Arina:2009uq,D'Eramo:2010ep,D'Eramo:2011ec,Belanger:2012vp,Aoki:2012ub}, especially in the presence of a matter-antimatter asymmetry \cite{D'Eramo:2011ec}.  In order to correctly compute the DM density, semi-annihilation must be included since it provides additional channels to dilute relic DM particles.  Recently, the impact of semi-annihilations was implemented in \texttt{micrOMEGAs} software package for the cases of $Z_3$ and $Z_4$ symmetries~\cite{Belanger:2012vp}.

\begin{figure}
\begin{center}
\subfloat[]{\label{fig:annvssemiannA}\includegraphics[scale=0.65]{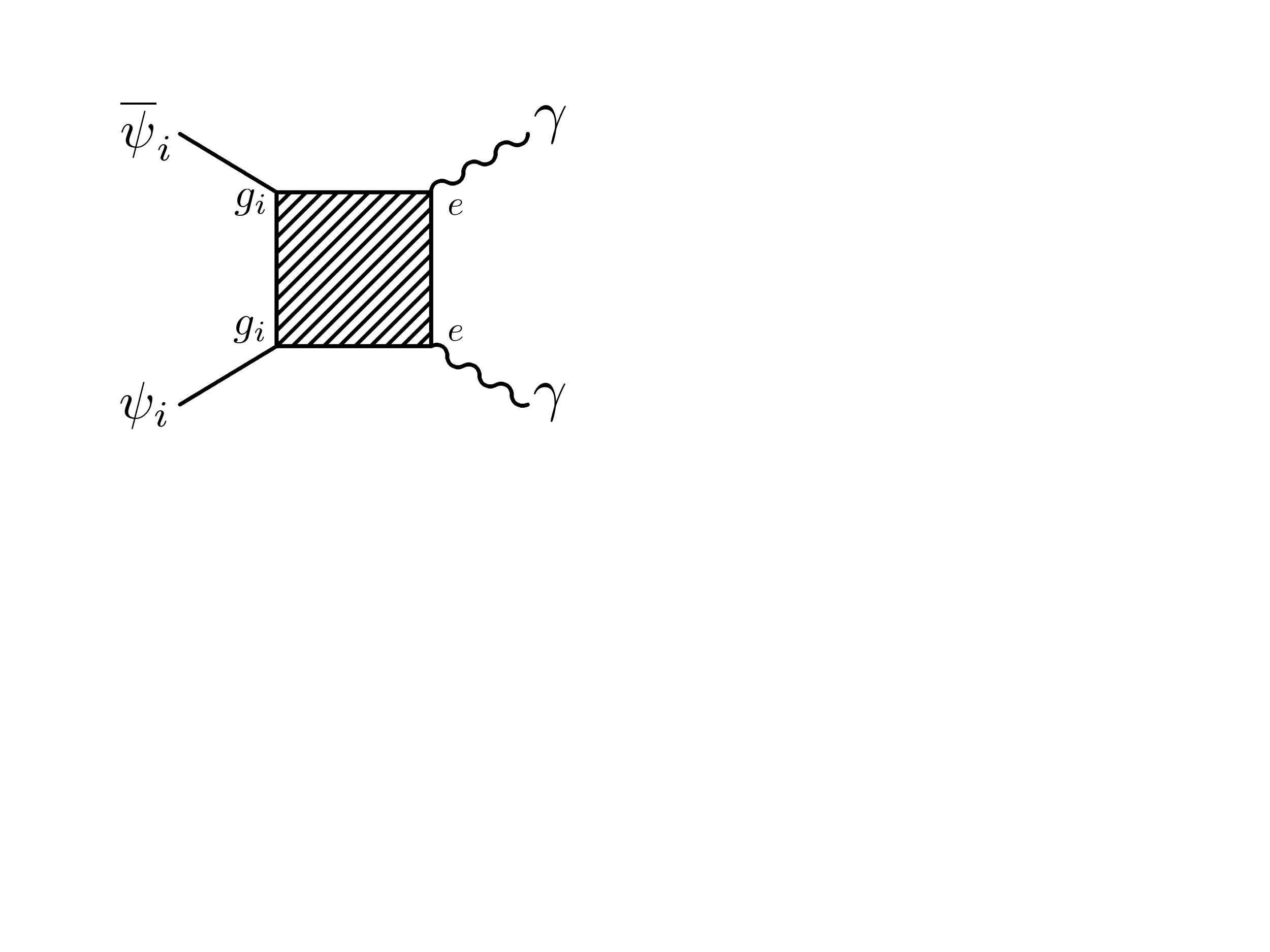}} $\qquad\qquad$
\subfloat[]{\label{fig:annvssemiannB}\includegraphics[scale=0.65]{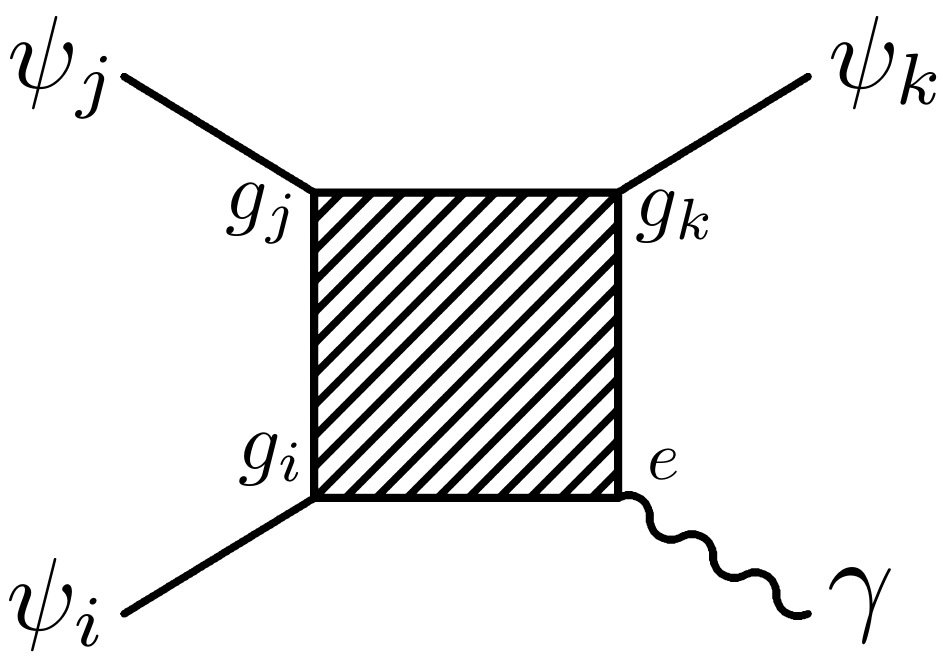}}
\caption{Schematic diagrams for sources of gamma ray lines from DM (semi-)annihilation.  Loops of charged particles are depicted by the hatched boxes. We show: (a) ordinary annihilation, where the cross section scales as $\sigma(\overline{\psi}_i \psi_i \rightarrow \gamma \gamma) \propto \alphaEM^2 \alpha_i^2$; (b) semi-annihilation, with cross section scaling as $\sigma(\psi_i \psi_j \rightarrow \psi_k \gamma) \propto \alphaEM \alpha_i \alpha_j \alpha_k$.}
\label{fig:AnnAndSemiAnn}
\end{center}
\end{figure}

\subsection{Gamma Lines for Indirect Direction}

Semi-annihilation not only affects DM production in the early universe but also leaves an impact today in DM indirect detection experiments.  The differential energy flux of $\phi$ particles from DM (semi-)annihilation in the Milky Way is of the form
\be
\frac{d \Phi_{\phi}}{d E_{\phi}} \propto \sum_{ij\rightarrow k} N_{\phi}^{ij \to k}\, n_i n_j \,\langle \sigma v\rangle_{ij \to k}\, \delta\left(E_{\phi} - E^{ij\to k}_{\phi}\right).
\label{eq:phispectrum}
\ee
Here, the indices $i$ and $j$ run over the various different DM species with number density $n_i$, and $N_{\phi}^{ij \to k}$ ($E^{ij\to k}_{\phi}$) are the number (energy) of $\phi$ particles produced in each reaction with averaged cross section $\langle \sigma v\rangle_{ij \to k}$.  For non-relativistic DM in the initial state, the spectrum of $\phi$ particles consists of several monochromatic lines with different intensities.  The spectrum of observed SM particles in indirect detection experiments depends on the nature of $\phi$.  If the field $\phi$ is unstable, we must to convolve the $\phi$ spectrum in \Eq{eq:phispectrum} with the decays of $\phi$.  Here, we focus on models where (semi-)annihilation directly produces gamma rays, namely $\phi = \gamma$.\footnote{For the case of semi-annihilation resulting in a neutrino final state, $\phi = \nu$, see \Ref{Aoki:2012ub}.} 

It is well-known that monochromatic photons can be produced in the DM annihilation process $\psi_i \overline{\psi}_i \to \gamma \gamma$ with energy
\be
\label{eq:lineann}
E^{i}_\gamma = m_i.
\ee
A schematic diagram for this process is sketched in \Fig{fig:annvssemiannA}.  Given the very stringent limits on charged relic DM \cite{McDermott:2010pa,Cline:2012is}, photons cannot be produced at tree-level through renormalizable DM-photon couplings.  Instead, annihilation proceeds via loop effects involving charged particles (or corresponding higher-dimensional operators).  Thus the cross section is expected to be proportional to $\alphaEM^2$, which is why standard DM scenarios predict a feeble direct annihilation rate to photons.  

In contrast, gamma ray production from semi-annihilation scales with only one power of $\alphaEM$, as depicted in \Fig{fig:annvssemiannB}.  Moreover, dark sector couplings can easily exceed the electromagnetic coupling while still remaining perturbative, so gamma ray production from semi-annihilations can be enhanced with respect to ordinary annihilations.

As mentioned in the introduction, the semi-annihilation reaction $\psi_i \psi_j \rightarrow \psi_k \gamma$ leads to gamma lines at energies
\be
\label{eq:linesemi}
E^{ij \to k}_\gamma = \frac{(m_i + m_j + m_k) (m_i + m_j - m_k)}{2 (m_i + m_j)},
\ee
enriching the spectrum of gamma rays expected in DM indirect detection.  In the simple case where all $m_i$ are equal, there are two lines at
\be
E^\psi_\gamma = m_\psi \quad (\text{annihilation}), \qquad E^{\psi\psi\psi}_\gamma = \frac{3}{4} m_\psi  \quad (\text{semi-annihilation}).
\ee
More general multi-component DM models with $N$ species would yield $N$ annihilation lines with energies given by \Eq{eq:lineann}, and as many as $N (N-1) (N-2)/2$ semi-annihilation lines with energies from \Eq{eq:linesemi}.  Note that the resulting gamma ray spectrum from annihilations are directly related to the DM mass.  On the other hand, the semi-annihilation spectrum is determined from the DM mass \emph{differences}, and one could explain a $130~\GeV$ line with all DM species in the multi-TeV mass range, or greater.  Of course, the specific intensities of the lines will depend on the detailed couplings and symmetries of the DM scenario.

\subsection{Direct Detection and Collider Phenomenology}

For a given indirect detection signal, one must consult a specific model to determine the possible collider and direct detection implications.  That said, we can make a few generic remarks about how semi-annihilation might modify DM phenomenology compared to ordinary annihilation.  Usually, when one imposes the requirement of a particular annihilation cross section, either to explain the DM relic abundance or an indirect detection signal, it is possible to use the reversed Feynman diagram to predict collider production rates and signatures.  Similarly, the same Feynman diagrams read on their side can be used to predict direct detection signals.  

In contrast, if the cosmology of DM is determined via semi-annihilation, this phenomenology is greatly changed.  Collider production via the semi-annihilation diagrams would result in three DM particles in the final state, leading to a much smaller production cross section and modified kinematics.  Similarly, direct detection via these diagrams would be quite suppressed, since elastic scattering diagrams only appear via DM loops.  Of course, generic semi-annihilation scenarios often allow for standard pair annihilation, though from the parametrics of \Fig{fig:AnnAndSemiAnn}, such annihilation diagrams are typically suppressed.  For these reasons, one generically expects potential collider or direct detection signals from a semi-annihilation scenario to be suppressed compared to a corresponding annihilation scenario.

\pagebreak

\begin{figure}
\centering
\includegraphics[height=1.6in]{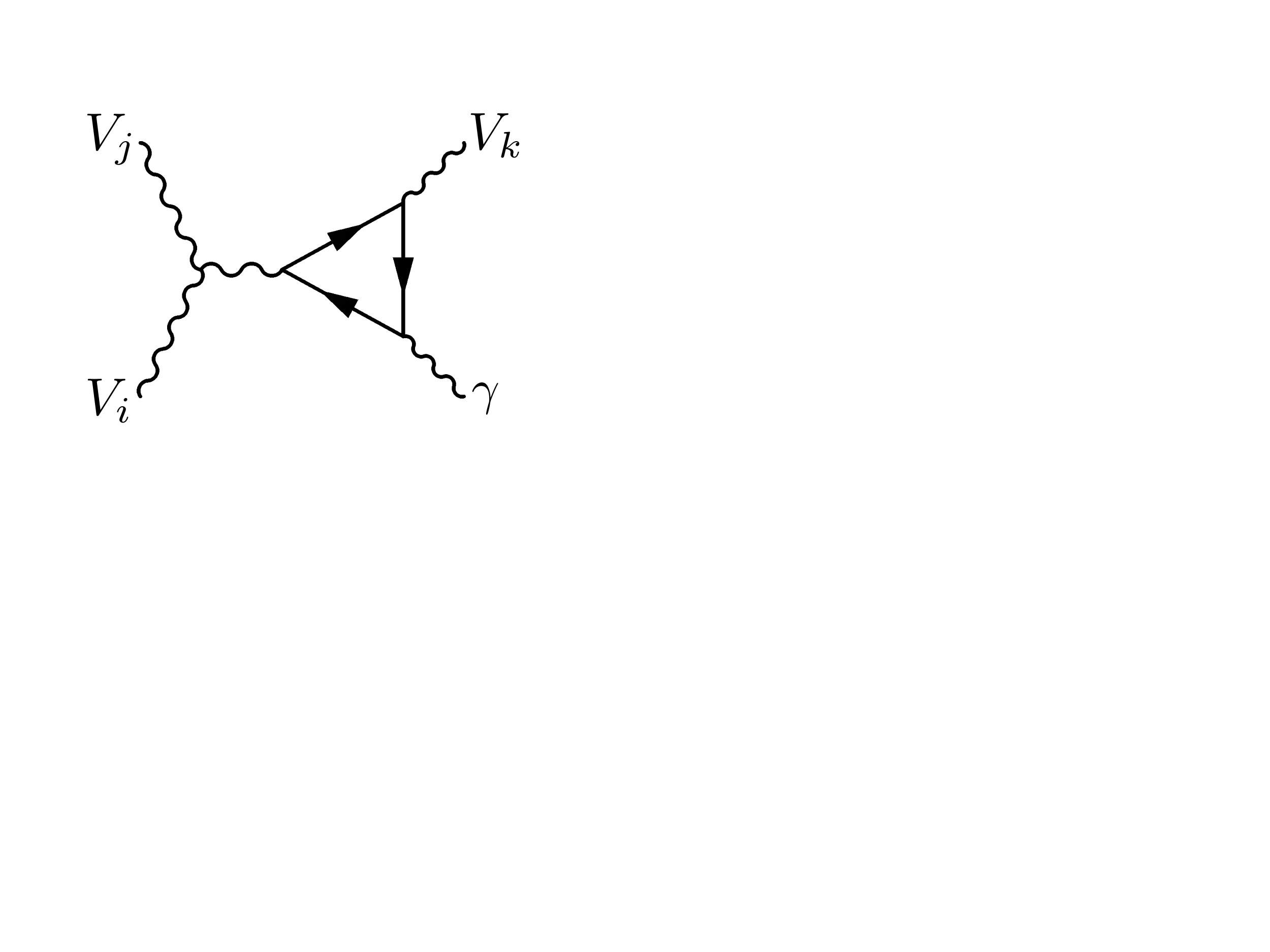} 
\caption{Gamma rays from semi-annihilation of vector DM through triangle diagrams.  In general such processes are allowed, however if the vectors originate from a spontaneously broken gauge group and the charged fermions are vector-like under all symmetries, these diagrams are forbidden by anomaly cancelation.}
\label{fig:semianntri}
\end{figure}

\section{Semi-Annihilation of Vector Dark Matter}
\label{sec:model}
We now construct an explicit class of models which give rise to gamma rays from semi-annihilation.  Considering \Fig{fig:annvssemiannB}, one could associate various spin-representations to $\psi_i$, $\psi_j$, and $\psi_k$.  Particle assignments could include a pair of fermions and a boson, or three bosons.  In supersymmetric scenarios, one can have particles of different spin transforming under a common DM stabilization symmetry.  For simplicity, we restrict DM to consist of particles of the same spin (ruling out the participation of fermions), though we consider models involving fermions in \Sec{sec:retrofit}.  In this section, we focus on massive spin-1 vectors as our DM candidates.  After introducing our vector DM model in \Secs{sec:genericbox}{sec:darkSU3}, we discuss additional direct detection and relic abundance considerations in \Secs{sec:messagerdirect}{sec:relic}.

\subsection{Gamma Rays from Box Diagrams}
\label{sec:genericbox}

Vector DM can arise in a number of ways \cite{Servant:2002aq,Cheng:2002ej,Hooper:2004xn,Cheng:2003ju,Birkedal:2006fz,Hambye:2008bq,Hambye:2009fg,Zhang:2009dd,Arina:2009uq,DiazCruz:2010dc,Bhattacharya:2011tr,Lebedev:2011iq,Farzan:2012hh}, including as resonances of some strongly-coupled dark sector \cite{Hambye:2009fg}.  Here we focus on massive gauge bosons from a spontaneously broken gauge group $\Gdark$, which are stabilized by a remnant custodial symmetry \cite{Hambye:2008bq}.  For simplicity, we assume that the vector DM particles share a common mass $M_V$ enforced by the custodial symmetry, though more general scenarios are possible.

\begin{figure}
\centering
\subfloat[]{\label{fig:vectorsemiann}\includegraphics[height=1.6in]{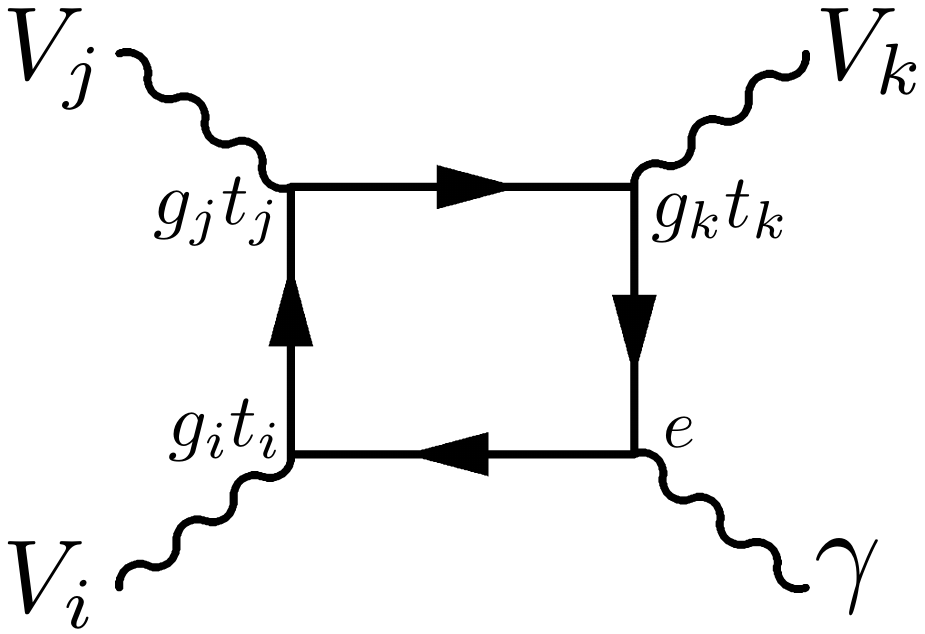}}   
\hspace{0.5in}
\subfloat[]{\label{fig:vectorann}\includegraphics[height=1.6in]{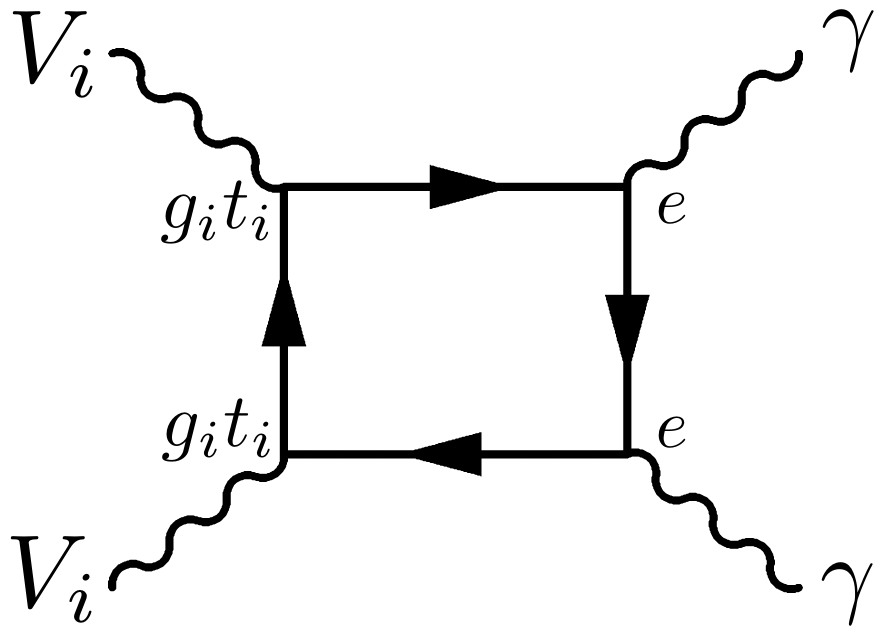}}
\caption{Production of gamma-ray lines from (a) semi-annihilation and (b) annihilation of dark gauge bosons for general dark gauge groups.  Here, $t^i$ are the generators and different gauge couplings are allowed for product gauge groups.  Permutations of external legs are not shown.}
\end{figure}

In order to generate gamma rays from semi-annihilation, we assume the existence of ``messengers'', namely matter which is charged both under electromagnetism and $\Gdark$.  We take the messengers to be fermions which carry vector-like charges under any symmetry.  We also assume that the messenger masses $M_M$ do not receive contributions from the higgsing of $\Gdark$.  As a consequence, for a spontaneously broken gauge group, the requirement of anomaly cancellation forbids semi-annihilation through triangle diagrams such as \Fig{fig:semianntri}.

This leaves only the box diagrams of \Fig{fig:vectorsemiann} as the origin of DM semi-annihilation.  These diagrams have a kinematic structure analogous to the light-by-light scattering in QED \cite{Euler:1935zz,Heisenberg:1935qt,Schwinger:1951nm}.  The calculation of semi-annihilation through box diagrams is somewhat more involved than light-by-light scattering since three of the external bosons are massive.   That said, general expressions for the QED box diagram were calculated for off-shell external photons by Karplus and Neuman in \Refs{Karplus:1950zza,Karplus:1950zz}, and these results can be employed for semi-annihilation.  Extending this calculation to the non-Abelian case is straightforward.  To lowest order in the messenger mass, the effective operator which is generated is a non-Abelian extension of the Euler-Heisenberg Lagrangian
\be
\mathcal{L} \supset  \sum_{24 \text{ perm}} \frac{g_i g_j g_k e}{180 (4 \pi)^2 M_M^4} \Tr [ t^i t^j t^k] \left(5 G^i_{\mu\nu} G^{j \nu\mu} G^k_{\lambda\rho} F^{\rho\lambda} - 14 G^i_{\mu\nu} G^{j \nu\lambda} G^k_{\lambda\rho} F^{\rho\mu} \right) ~~,
\ee
where the sum is over all possible $24$ permutations of the four field strengths.\footnote{In the general case, the appropriate combinatorial factors should be included for identical fields, as in for photon-photon scattering.  This is not a concern here, though, since the amplitude vanishes for any identical non-Abelian vectors.}  This lowest-order effective coupling is sufficient for illustrative purposes, however we have calculated corrections at all orders in $M_V/M_M$ by evaluating the Feynman parameter expressions found in \Refs{Karplus:1950zza,Karplus:1950zz}.

\begin{figure}
\centering
\includegraphics[height=2.7in]{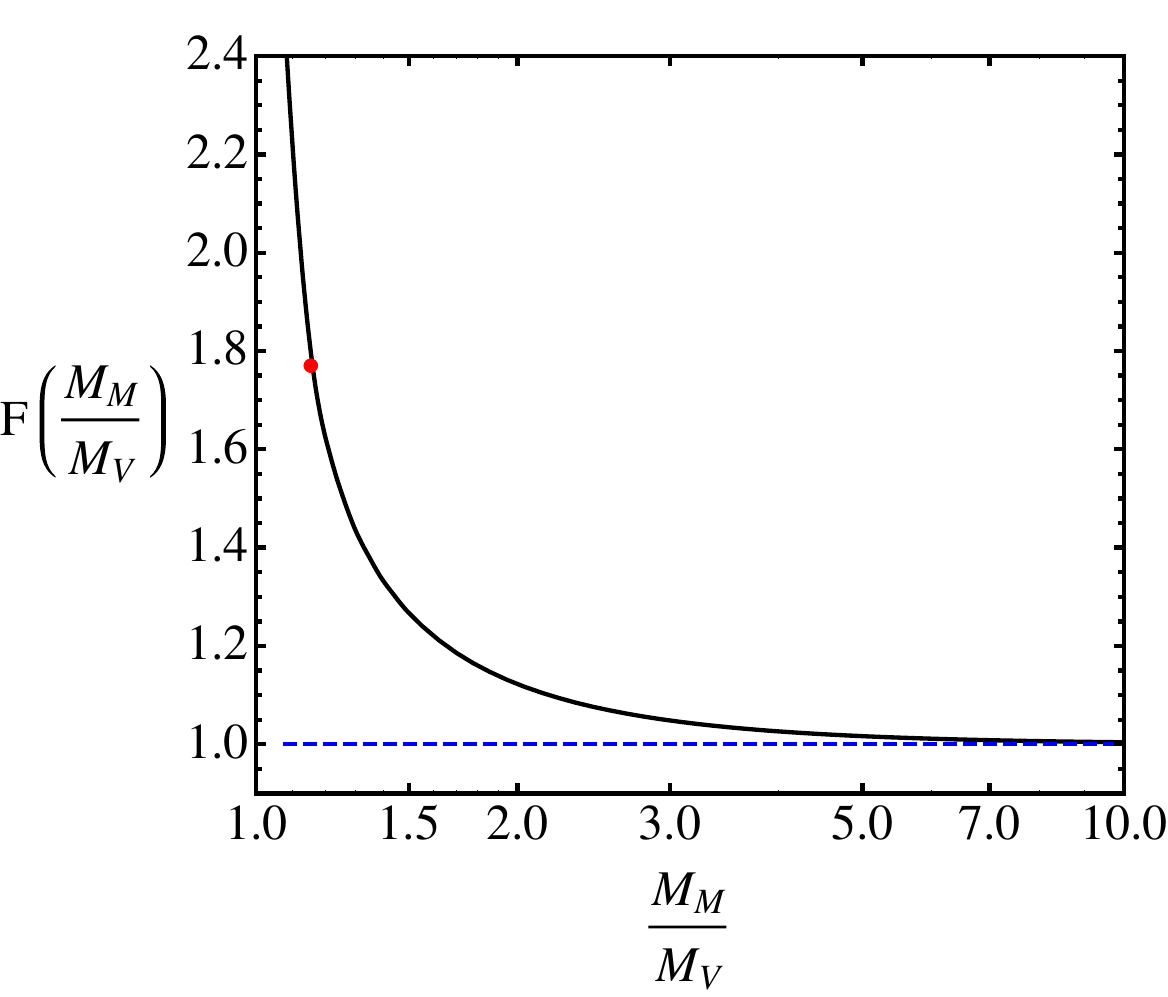} 
\caption{Enhancement of the full loop integral in comparison with the lowest-order terms found using the Euler-Heisenberg Lagrangian.  For large messenger masses, the form factor approaches unity (dashed blue) and the Euler-Heisenberg Lagrangian is sufficient. As the messenger masses approach the vector masses, the higher-order corrections become important.  The red dot corresponds to $M_M/M_V = 200/173$, which is the benchmark point used for the $\SU(3)_d$ model of \Sec{sec:darkSU3}.}
\label{fig:numerical}
\end{figure}

We will always assume that the messengers are more massive than the DM, i.e.\ $M_V<M_M$, forbidding direct DM annihilation into the charged fermions.  Since the leading order term in the semi-annihilation amplitude scales as $\mathcal{M} \sim (M_V/M_M)^4$, we will require that the messengers are not too heavy.  In the simplifying limit in which $M_i = M_j = M_k = M_V$ and taking all dark gauge couplings to be equal,\footnote{The extension to product gauge groups is straightforward.}  the final result for $s$-wave semi-annihilation $\vev{\sigma v}_s$ of species $i$ and $j$ into species $k$ and a photon is
\be
\vev{\sigma v}_s (i j\rightarrow k \gamma) = \frac{1697}{460800\pi} \frac{\alphaD^3 \alphaEM}{M_V^2} \left( \frac{M_V}{M_M} \right)^8  |T^{ijk}|^2  F \left( \frac{M_M}{M_V} \right) ~~.
\label{eq:genresult}
\ee
Here, $F(M_M/M_V)$ is a form factor which parameterizes the deviation from the leading order result in an expansion in the vector to messenger mass ratio.  Closed-form expressions in terms of Spence functions can be found by reducing the loop integral to the usual basis set of scalar loop integrals, however the final expression is extremely lengthly. We have calculated this form factor numerically and the result is plotted in \Fig{fig:numerical}.  It is clear that for $M_M \gtrsim M_V$, enhancements of $\mathcal{O} (1)$ arise.

The group-theoretic factor in \Eq{eq:genresult} is
\be
T^{ijk} = \sum_t \Tr[t^i t^j t^k + t^i t^j t^k] ,
\ee
where the sum runs over the vector-like messengers transforming in the $t^i$ representation of $\Gdark$.\footnote{Because the sum is over vector-like fermions, anomaly cancellation does \emph{not} enforce $T^{ijk} = 0$.  Similarly, imposing a charge conjugation symmetry would not imply $T^{ijk} = 0$.}  Note that $T^{ijk}$ scales like $N_f$, where $N_f$ is the number of messengers, so $\vev{\sigma v}_s$ scales like $N_f^2$.  The total DM semi-annihilation cross section relevant to indirect detection is
\be
\label{eq:totalsemi}
\vev{\sigma v}_s (V V\rightarrow V \gamma) = \sum_{i,j,k} f_i \, f_j\, \vev{\sigma v}_s (i j\rightarrow k \gamma),
\ee
where $f_i$ is the fraction of DM made up by species $i$.  With an unbroken custodial symmetry, $f_i = 1/N_V$ where $N_V$ is the number of vector DM species.

\subsection{The Case of a Dark $\SU(3)$}
\label{sec:darkSU3}

To study the semi-annihilation process concretely, we now commit to a particular dark gauge group.  While a $\U(1)_d$ gauge group could exhibit semi-annihilation, we know of no symmetry that would allow semi-annihilation without also allowing kinetic mixing between the dark $\U(1)_d$ and $\U(1)_Y$, rendering the vector DM unstable.  The case of $\SU(2)_d$ also does not work because $\SU(2)_d$ has (pseudo-)real representations and thus $T^{ijk} = 0$.\footnote{One could consider the product dark group $\SU(2)_d \times \U(1)_d$, which has some elements with $T^{ijk} \not= 0$.}  The smallest $\SU(N)_d$ group for which semi-annihilation can be achieved is $\SU(3)_d$, and we will focus on this case.  

If the dark $\SU(3)_d$ is broken by a $3\times3$ bi-fundamental Higgs with vacuum expectation value proportional to the identity,\footnote{Such a spontaneous breaking pattern can be enforced with an appropriate flavor symmetry.} then all of the $\SU(3)_d$ gauge bosons obtain the same mass, and are stable owing to the remaining custodial symmetry.  In this case, all eight gauge bosons form equal components of the DM and each species makes up a fraction $f=1/8$ of the DM density.  Since the vectors all have the same mass, the vector masses must be $M_V \simeq 173~\GeV$ to produce a gamma line at $130~\GeV$.  We will take $N_f$ messengers in the fundamental of $\SU(3)_d$ carrying electric charge $1$,\footnote{We will discuss the possible $\SU(2)_L \times \U(1)_Y$ quantum numbers of the messengers in the next subsection.} yielding the averaged group-theoretic factor
\be
\label{eq:traceSU3}
\sum_{i,j,k} \frac{1}{f_i} \frac{1}{f_j} \left|T^{ijk} \right|^2 = \frac{5}{96} N_f^2.
\ee

\begin{figure}
\centering
\includegraphics[height=2.6in]{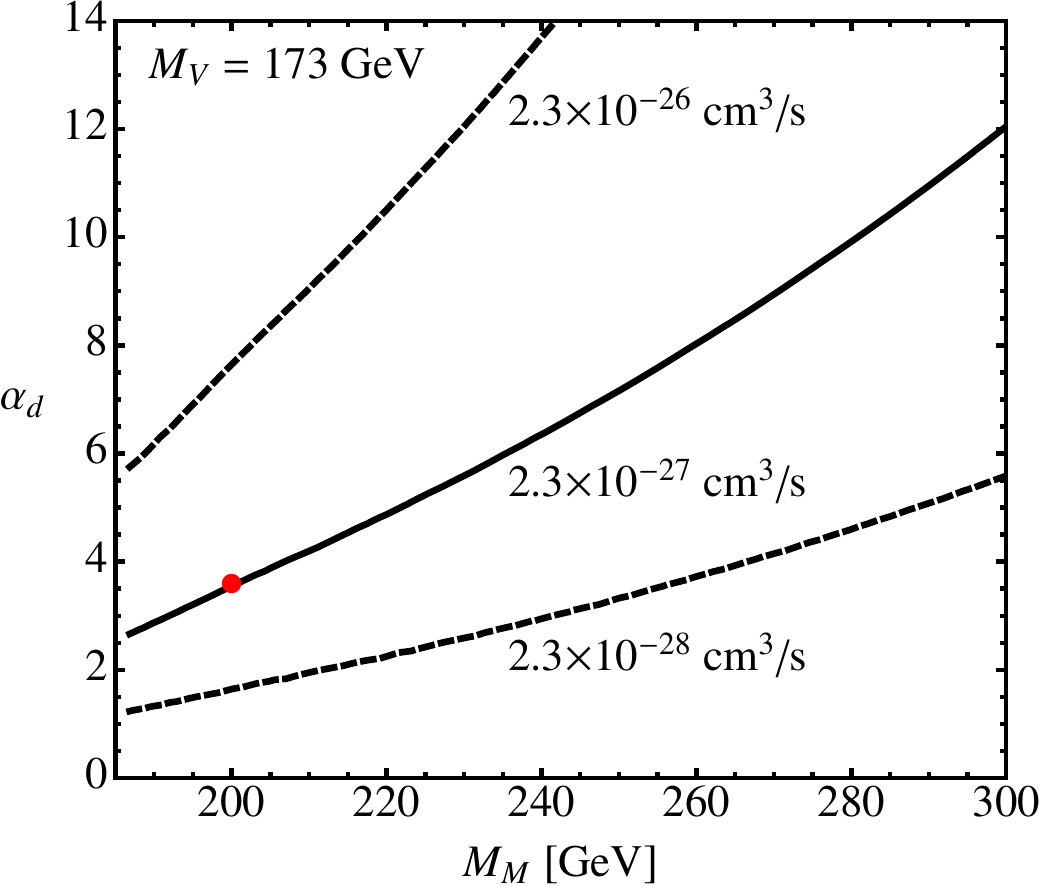} 
\caption{Contours in the  $M_M-\alpha_d$ plane of constant $\frac{1}{2} \vev{\sigma v}_s (V V\rightarrow V \gamma)$ for the $\SU(3)_d$ model with $M_V = 173$ GeV.  These are given in units of $2.3 \times10^{-27} \, \text{cm}^3/\text{sec}$, which is the approximate cross section required to explain the \Fermi line after the DM density is adjusted to account for the increased DM mass.  The form factor $F(M_M/M_V)$ from \Fig{fig:numerical} is included.  For messenger masses $M_M \gtrsim M_V$ perturbative values of the dark gauge coupling are sufficient, and for higher masses one can see the expected $\alpha_d \propto M_M^{8/3}$ scaling of the required coupling.}
\label{fig:numericalsigv}
\end{figure}

Summing over all species in \Eq{eq:totalsemi}, the full $s$-wave semi-annihilation cross section is
\begin{eqnarray}
\frac{1}{2} \vev{\sigma v}_s (V V\rightarrow V \gamma)  & =  & \frac{5}{192} \frac{1697}{460800\pi} \frac{ \alphaD^3 \alpha}{M_V^2} N_f^2 \left( \frac{M_V}{M_M} \right)^8 F \left( \frac{M_M}{M_V} \right)  \label{eq:semisu3} \\
& \simeq & 2.3 \times 10^{-27} \, \text{cm}^3  / \text{s} \left( \frac{\alphaD}{3.55} \right)^3 N_f^2 \left( \frac{200~\GeV}{M_M} \right)^8 \left( \frac{M_V}{173~\GeV} \right)^6, \nonumber
\end{eqnarray}
where in the second line we have used the numerical value of the form factor for this choice of masses, and the cross section is divided by two since we only observe one photon per annihilation.  This benchmark cross section is a factor of $(173/130)^2$ larger than the nominal $1.3\times10^{-27} \, \text{cm}^3  / \text{s}$ value required for annihilating DM \cite{Weniger:2012tx} to counteract the smaller number density for larger DM masses.  The reference values were chosen to allow the desired semi-annihilation rate to explain the \Fermi line using a quasi-perturbative value of $\alphaD \lesssim 4 \pi$, without requiring too much tuning between the messenger mass and the DM mass.  The reference value $N_f = 1$ predicts the existence of three ($=3N_f$) new charged fermions present just above the weak scale.  Contours of $\vev{\sigma v}_s (V V\rightarrow V \gamma)$ in the $M_M - \alpha_d$ plane are shown in \Fig{fig:numericalsigv}, where the full form factor dependence on the messenger mass is included.  

Note that the rate for semi-annihilation is somewhat smaller than one would naively estimate from the diagram in \Fig{fig:vectorsemiann}.  The reason for this is twofold.  First, the box diagram simply has a numerically small coefficient, beyond the estimate from naive loop counting.  Second, there is a suppression coming from the large number of DM species ($N_V = 8$).  Since there are only a small number of non-vanishing traces for semi-annihilation in \Eq{eq:traceSU3}, this leads to a fractional suppression of the effective cross section by $1/N_V^2$.  Despite these suppressions, \Eq{eq:semisu3} show that semi-annihilation can explain the \Fermi line for a perturbative value of $\alphaD$.

In addition to semi-annihilation, one has ordinary pair annihilation $V V \rightarrow \gamma \gamma$ shown in \Fig{fig:vectorann}.  This yields a higher-energy gamma line at $E_\gamma \simeq 173~\GeV$ and two photons per annihilation, with a cross section given by
\begin{eqnarray}
\vev{\sigma v}_s (V V\rightarrow \gamma \gamma)  & \approx  & \frac{1}{8} \frac{299 \alphaD^2 \alpha^2}{36450\pi M_V^2}  N_f^2 \left( \frac{M_V}{M_M} \right)^8 \label{eq:vecpairann} \\
& \simeq & 3.0 \times 10^{-29} \, \text{cm}^3  / \text{s} \left( \frac{\alphaD}{3.55} \right)^2 N_f^2 \left( \frac{200~\GeV}{M_M} \right)^8 \left( \frac{M_V}{173~\GeV} \right)^6, \nonumber
\end{eqnarray}
where we have not included the form factor enhancement in this estimate.  At present, a $173~\GeV$ line at this cross section is below current \Fermi sensitivity, but would be a smoking gun for the semi-annihilation scenario if present.  Finally, annihilation into $\gamma Z$ is also possible, though the detailed rate depends on the electroweak charges of the messengers.  Generically, the $\gamma Z$ rate will be comparable to the $\gamma \gamma$ rate, yielding an additional gamma line at $E_\gamma \simeq 161$ GeV.

It is clear that explaining the \Fermi line via the semi-annihilation box diagram does require a large dark gauge coupling, with increasing values of $\alphaD$ needed for heavier messenger masses.  On the other hand, \Eq{eq:vecpairann} shows that an explanation of the \Fermi line based on an ordinary annihilation box diagram would require an even larger gauge coupling (or a larger number of messengers).  In this way, semi-annihilation processes can be enhanced compared to annihilation processes, making them a plausible DM explanation for the \Fermi line.  We will explore other ways that semi-annihilation can dominate over annihilation in \Sec{sec:retrofit}.

\subsection{Messenger Direct Detection and Collider Bounds}
\label{sec:messagerdirect}

We have seen that $\SU(3)_d$ vector DM can yield a gamma line via semi-annihilation.  Here, we consider possible constraints on such a scenario from direct detection and collider experiments.  In the next subsection, we discuss relic abundance constraints.

Without introducing additional couplings between the messengers and the visible sector, the messengers themselves are stable and thus contribute to the observed DM density.  In addition, the messengers must consist of electroweak multiplets containing both charged and neutral fields, since if they only contained charged fields the model would face strong bounds on charged relics \cite{McDermott:2010pa,Cline:2012is}.  For typical electroweak charges, radiative corrections split the degeneracy between the charged and neutral components, leading to a mass splitting $M_{\pm}-M_0 \sim \mathcal{O} (100)$ MeV \cite{Cirelli:2005uq}.  For this reason, charged relics are of no concern, since the charged fields will decay down to the neutral fields via virtual $W$ boson emission.

However, the neutral relics can still be a concern because of stringent direct detection constraints.  For Dirac fermion messengers, the observed DM abundance is saturated for $M_M \sim$ few TeV \cite{Cirelli:2005uq}.   Since the abundance scales as $\Omega_M h^2 \propto M_M^2$, then for $M_M \sim 200$ GeV the messengers will comprise only an $\mathcal{O} (1/100)$ fraction of the observed DM abundance.  While this fraction is small enough not to affect the \Fermi line, it is large enough to face direct detection bounds.  Messengers with non-zero hypercharge experience spin-independent elastic scattering on nuclei mediated via $Z$-boson exchange, with a cross section well above current bounds even with a 1/100 dilution of the abundance.  For this reason, we opt for $\SU(2)_L$ triplet messengers with zero hypercharge, since spin-independent messenger-nucleon scattering via $Z$-boson exchange is forbidden.\footnote{Alternatively, we could mix the Dirac $\SU(2)_L$ doublets with a Majorana singlet to render the spin-independent process inelastic.}  In this case, a spin-independent cross section $\sigma \sim 10^{-45}~\text{cm}^2$ is generated at the one-loop level \cite{Cirelli:2005uq}, which is well below current bounds when combined with the low messenger abundance.

Stable $\SU(2)_L$ triplet messengers also have an interesting collider phenomenology, similar to nearly pure winos in supersymmetric scenarios.  The charged messengers can be pair produced weakly via the Drell-Yan process and then decay within a few centimeters to the neutral partner and a very soft pion.  Such processes are extremely difficult to observe, however promising approaches have been demonstrated \cite{Buckley:2009kv}.  Detection of these messengers typically requires large integrated luminosities, and messengers with mass $M_M \approx 200$ GeV are consistent with current LHC bounds \cite{ATLAS:2012ab}.  Depending on the lifetime of the charged messengers, potentially stronger bounds are provided by \Ref{Chatrchyan:2012sp}.

Of course, the main phenomenological feature of this work, namely gamma lines from DM semi-annihilation, does not rely on a specific model for messengers, and we have outlined only one possibility here.  Similarly, if one allows for additional fields and couplings, then the messengers could be made unstable,\footnote{As an example, messengers with non-trivial lepton number could decay to leptons and weak bosons.} removing any relic messengers and greatly changing the collider phenomenology.

\subsection{Relic Abundance}
\label{sec:relic}

A key question for any WIMP DM scenario is whether the DM relic abundance could be determined via thermal freeze out.  As currently presented, the interactions in \Sec{sec:darkSU3} are too feeble to yield the correct relic density.  The loop-level (semi-)annihilation cross sections are too small to sufficiently dilute DM, and since $M_M > M_V$, tree-level annihilation to messengers is kinematically inaccessible at the desired freeze out temperature $T_f \simeq M_V / 20$.  

There are a number of avenues to achieve the correct relic density.  First, one can keep the current field content and lower the mass of the messengers such that $M_V$ and $M_M$ are nearly degenerate.  In this scenario, DM freeze out tracks messenger freeze out until late times.  This possibility has been discussed in \Ref{Tulin:2012uq} and can lead to the correct relic abundance via a form of assisted freeze out \cite{Belanger:2011ww}.   However, this requires a close coincidence between messenger and DM masses.  

\begin{figure}
\centering
\includegraphics[height=1.0in]{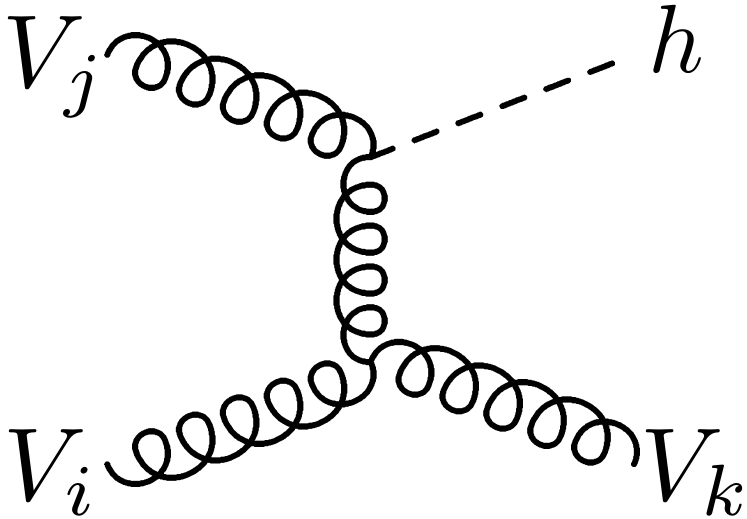}  \hspace{0.5in} \includegraphics[height=1.0in]{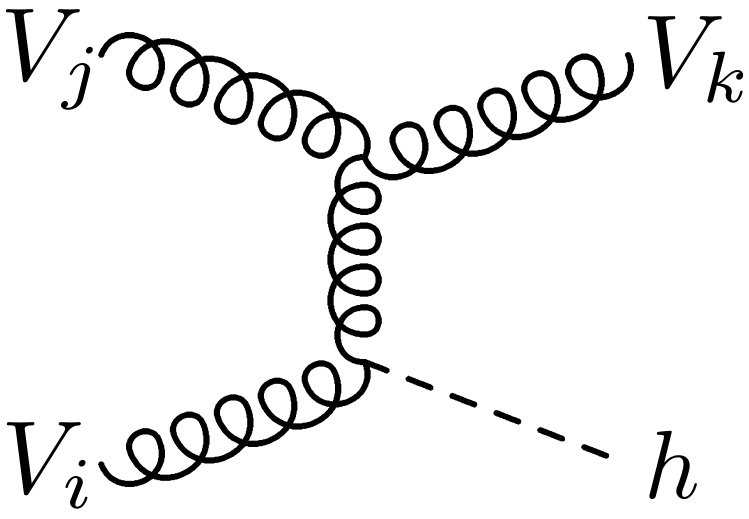}      \hspace{0.5in} \includegraphics[height=1.0in]{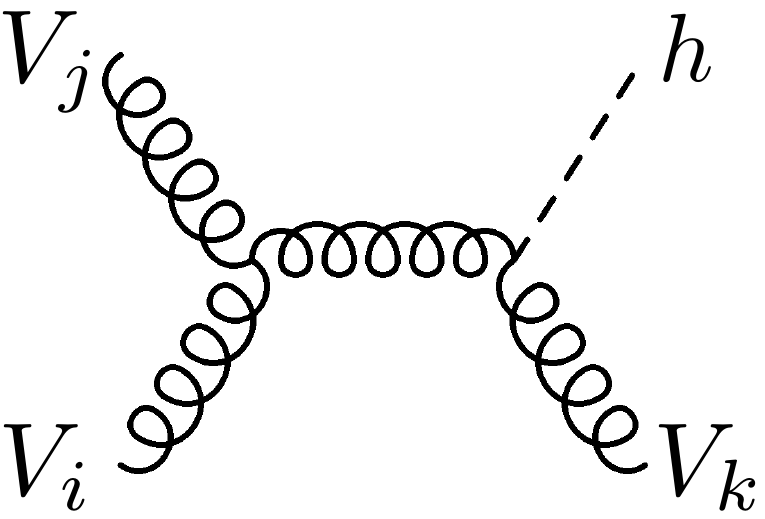}    
\caption{Tree-level semi-annihilation of vector DM.  This cross section scales with the mixing angle as $\vev{\sigma v}_s (VV\rightarrow V h) \propto \sin^2 \theta_h$ and can dominate freeze out even for small mixing angles.}
\label{fig:vectorhiggsann2}
\end{figure}

A simple alternative is to employ the Higgs portal \cite{Silveira:1985rk,McDonald:1993ex,Burgess:2000yq,Davoudiasl:2004be,Patt:2006fw}.  The fields which break the dark force are three fundamentals of $\SU(3)_d$, and can be written as a $3 \times 3$ matrix of fields $\Hdark$.  These fields can couple to the visible sector through the Higgs portal coupling $\mathcal{L} \supset \lambda |\Hdark|^2 |H|^2$, which respects the custodial symmetry.  These dark Higgses obtain a vev $\langle \Hdark \rangle = v_d I_3/\sqrt{2}$ and, after electroweak symmetry breaking, the diagonal component of $\Hdark$ mixes with the SM Higgs $h_0$, introducing couplings between the dark vectors and the SM Higgs.  We can write these couplings as
\be
\label{eq:darkHiggsCoupling}
\mathcal{L} \supset \sum_{i} \frac{1}{2}  M_V^2 A^{i\mu} A^i_\mu \left(1+  \frac{\sin \theta_h h_0}{v_d} +...\right)^2,
\ee
where $\theta_h$ is the mixing angle between dark and SM Higgses, $M_V^2 = g_d^2 v_d^2/2$, and the ellipsis denote additional dark Higgs states.  Since the SM Higgs only mixes with the diagonal component of $\Hdark$ and the $\SU(3)_d$ generators are traceless, all cubic $Vhh$ couplings vanish.  This leaves the couplings $VVh$ and $VVhh$, which are diagonal in $\SU(3)_d$ since $\Tr[t^i t^j] = \delta^{ij}/2$.  

The interactions in \Eq{eq:darkHiggsCoupling} allow vector DM to annihilate into pairs of SM fields and to semi-annihilate into a vector and a Higgs.  Note that such processes are kinematically accessible because $M_V \simeq 173~\GeV$ and $m_h \simeq 125~\GeV$.  For simplicity, we will assume all other states (in particular the dark Higgs bosons) are heavier than the DM.

We will calculate semi-annihilation first, since this will turn out to be the dominant interaction.  The diagrams responsible for semi-annihilation are depicted in \Fig{fig:vectorhiggsann2}.  Taking into account the relative fractional abundance of each vector, and summing over all species, the $s$-wave DM semi-annihilation cross section is
\be
\vev{\sigma v}_s (VV\rightarrow V h)  =  \frac{9 \pi \alphaD^2 \sin^2 \theta_h}{32 M_V^2} \frac{\left( 1 -\frac{m_h^2}{9 M_V^2} \right)^{3/2} \left( 1 -\frac{m_h^2}{M_V^2} \right)^{3/2}}{\left( 1 -\frac{m_h^2}{3 M_V^2} \right)^{2}}  ~~.
\ee
Inputting $m_h = 125~\GeV$ and $M_V = 173~\GeV$ yields
\be
\frac{1}{2} \vev{\sigma v}_s (VV\rightarrow V h)  = 2.9 \times 10^{-26}  \, \text{cm}^3  / \text{s} \left( \frac{\alphaD}{3.55} \right)^2 \left( \frac{\sin \theta_h}{0.0055} \right)^2,
\ee
which is of the parametrically required size to explain the vector DM relic density \cite{D'Eramo:2010ep}.\footnote{Using the techniques of \Ref{D'Eramo:2010ep}, one can find a semi-analytic solution for $\Omega_V h^2$ using the standard thermal freeze out approximation with the modified Boltzmann equation.  Only one Boltzmann equation is needed, since the DM particles all have the same mass and carry the same DM number.   The factor of $1/2$ is important and can be understood as arising because each semi-annihilation only changes the total number of DM particles by one unit relative to annihilation.}   For the expected Higgs decay channels, such thermal freeze out cross sections are below, but approaching, current indirect detection bounds on continuum photons \cite{Ackermann:2011wa,Cohen:2012me,Buchmuller:2012rc},\footnote{Each semi-annihilation typically results in a $\overline{b} b$ pair, each with an average energy of $76$ GeV, and the relevant constraint can be roughly estimated from the constraints on $76$ GeV DM annihilating to $\overline{b} b$ pairs, with the appropriate re-scaling due to different DM number densities.} and below current bounds on positrons, anti-protons \cite{Buchmuller:2012rc}, neutrinos \cite{Cirelli:2012tf}, and radio emissions \cite{Laha:2012fg}.

\begin{table}[t]
\centering
\begin{tabular}{c || c}
Final State & Scaling \\
\hline \hline
$V h$ & $\alphaD^2 \sin^2 \theta_h$ \\
\hline
$h h$ & $\alphaD^2 \sin^4 \theta_h$\\
$h^* \to WW, ZZ$ & $\alphaD \alpha_W \sin^2 \theta_h$ \\
$h^* \to \overline{t} t$ & $\alphaD \alpha_t \sin^2 \theta_h \langle p^2 / M_V^2 \rangle$
\end{tabular}
\caption{Relative scaling of various DM annihilation final states.  Here, $h^*$ represents an $s$-channel off-shell Higgs boson.  With our benchmark values, the $Vh$ semi-annihilation process dominates the DM relic computation.}
\label{tab:annscaling}
\end{table}

Since the dark gauge coupling is large, only a small mixing angle is required to maintain thermal equilibrium with the SM.  The mixing angle must be less than $\sin \theta_h \lesssim 0.0055$, otherwise the vector DM abundance would be rapidly depleted.  With this in mind, one can estimate the relative importance of other possible annihilation channels through their scaling with the various couplings.  In \Tab{tab:annscaling}, we show the relative scaling for various annihilation final states.  Since the mixing angle is small, annihilations to Higgs pairs will give at most a few percent correction.  Similarly, annihilations to electroweak bosons through an $s$-channel Higgs is also suppressed since $\alphaD \gg \alpha_W$.  Annihilation to top quark pairs could lead to corrections as large as $10\%$ at high temperatures since $\alpha_t \approx 1/4 \pi$, however DM annihilations to top quarks is strongly phase space suppressed, leading to additional suppressions by a factor $T/M_V \sim 1/20$ at freeze out.  Hence in this scenario, DM freeze out is dominated by semi-annihilation into a SM Higgs.  

As a final note, the required mixing angle $\theta_h$ is consistent with fits to Higgs observation data and also limits on additional Higgs-like scalars \cite{Bertolini:2012gu}.  Interestingly, if the dark Higgs states have mass below $M_{h_d} \lesssim 600$ GeV then, due to mixing with the SM Higgs, they could show up at the LHC through their decays to $WW$ and $ZZ$ pairs.

\section{Retrofitting Existing Models: RayDM}
\label{sec:retrofit}
Going beyond the vector DM models already discussed, one could build models involving more general combinations of fermions, scalars, and vectors.  The number of such possibilities is quite large, and we will not attempt an exhaustive classification.  Here we focus on the interesting case of taking existing models that explain the $130~\GeV$ line via annihilation, and retrofitting them to yield gamma lines from semi-annihilation.  In particular, the large couplings that are typically required for annihilation can be reduced to more moderate values in semi-annihilation.  In addition, the DM masses can take a broader range of values and still yield the $130~\GeV$ feature.

As an example, we consider a variant of Rayleigh DM (RayDM) \cite{Weiner:2012cb,Weiner:2012gm}, where DM is either a Majorana or pseudo-Dirac fermion.  These papers showed how the \Fermi $130~\GeV$ line could be explained by DM with mass $M_{\chi} \sim 130~\GeV$ which annihilates through loops of messengers (electroweak charged scalars and fermions) to on-shell photons.  For simplicity, we will consider a single Majorana fermion, in which case the transition dipole operator vanishes and all annihilation is achieved via the CP-odd Rayleigh operator.

\begin{figure}
\centering
\subfloat[]{\label{fig:raydm}\includegraphics[height=1.4in]{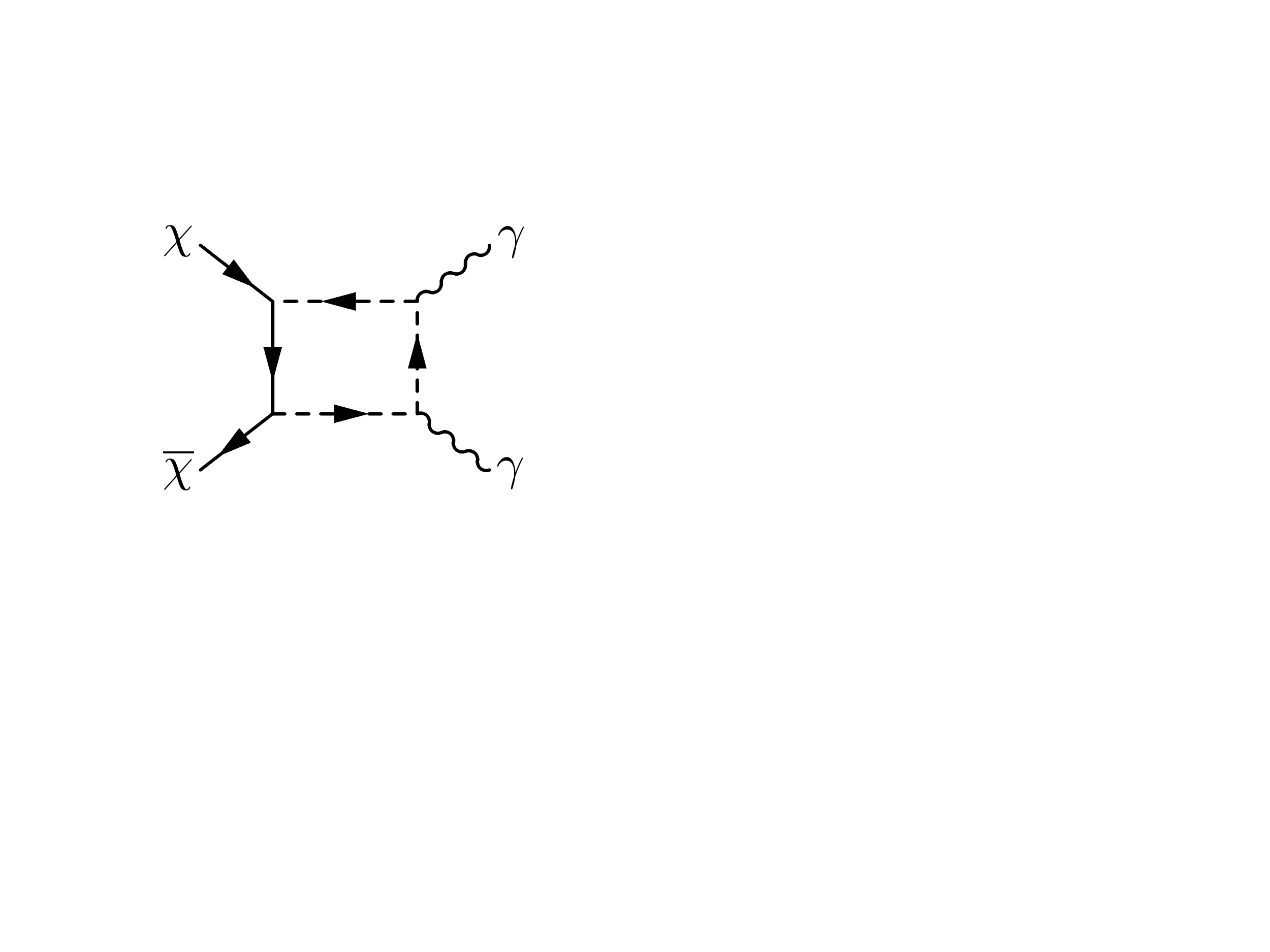}}   
\hspace{1.0in}
\subfloat[]{\label{fig:rraydm}\includegraphics[height=1.4in]{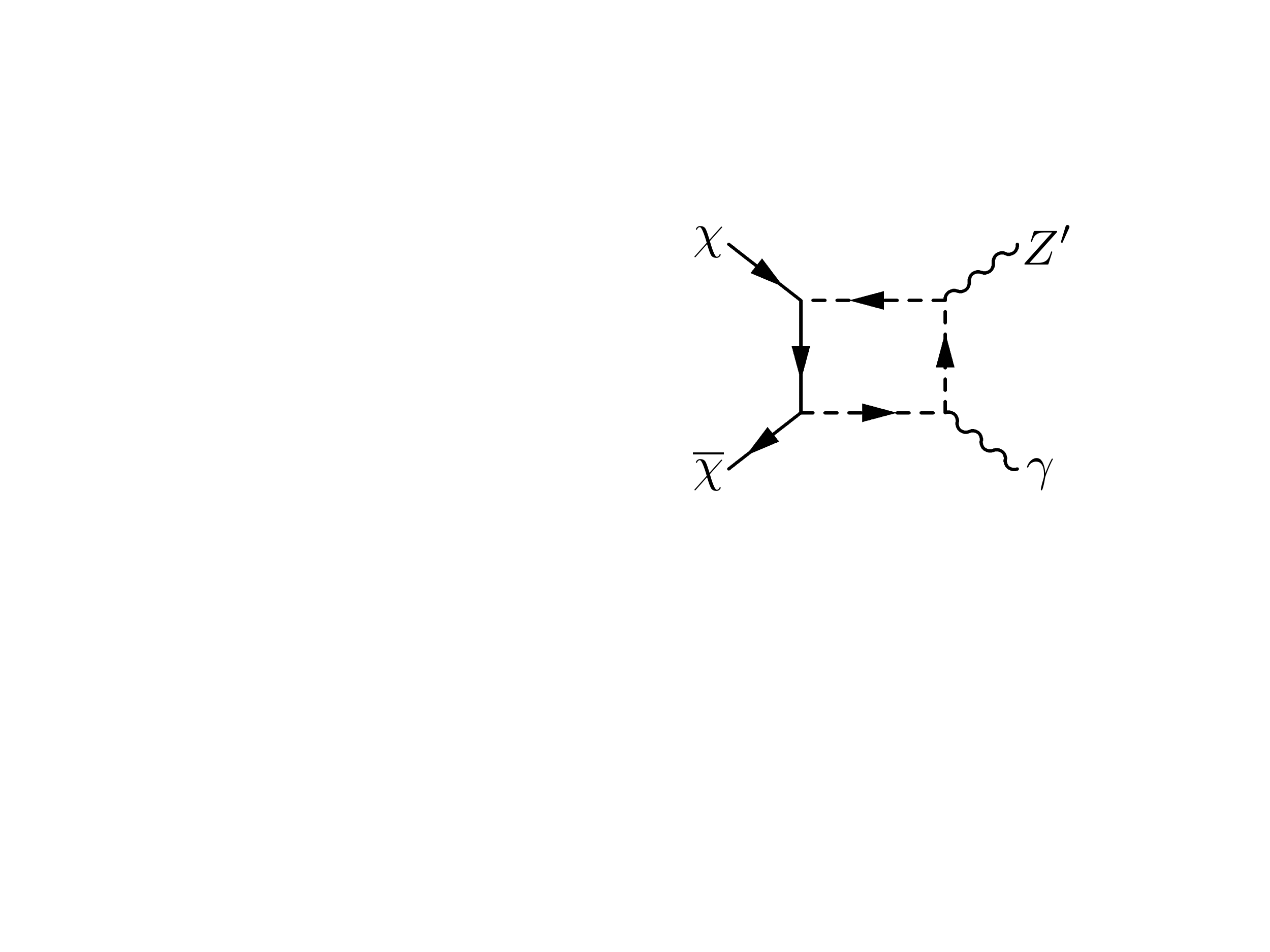}}
\caption{Typical diagrams contributing to (a) RayDM and (b) a retrofitted semi-annihilation version of RayDM with an additional dark sector final state $Z'$.  The loops involve both scalar and fermion messengers.}
\label{fig:rRayDM}
\end{figure}

To retrofit RayDM models to incorporate semi-annihilation, we simply introduce an additional $Z'$ which couples vectorially to the messengers.  RayDM and retrofitted RayDM scenarios are depicted in \Fig{fig:rRayDM}.  We choose this $Z'$ to have a mass $M_{\chi} < M_{Z'} < 2 M_{\chi}$, such that the additional process $\chi \chi \rightarrow \gamma Z'$ is allowed but $\chi \chi \rightarrow Z' Z'$ is kinematically forbidden.  In general, one expects the $Z'$ to kinetically mix with hypercharge and thus become unstable.  While the process $\chi \chi \rightarrow \gamma Z'$ would then technically be an annihilation diagram since it changes DM number by two units, we will continue to call it a semi-annihilation diagram since it has semi-annihilation kinematics (i.e.~$M_\chi$ and $M_{Z'}$ satisfy the triangle inequality in \Fig{fig:triangle}).  One could also envision semi-annihilation scenarios with a stable (or meta-stable) $Z'$-like state.

The electroweak charges of the messengers are free parameters and so we must parameterize this freedom.  Attempting to remain as close as possible to the conventions of \Refs{Weiner:2012cb,Weiner:2012gm}, we adopt the convention that, after integrating out the messengers, $\theta_{\chi}$ parameterizes the relative couplings of the DM to the hypercharge and $\SU(2)_L$ field strengths.  We then define the additional parameter $r'$ which gives the relative DM couplings to $\U(1)'$ and $\U(1)_Y$.  This yields the Rayleigh operators
\be
\mathcal{L} \supset \frac{\overline{\chi} \gamma_5 \chi}{4 \Lambda^3_R} \left( \cos \theta_\chi (B_{\mu \nu} \widetilde{B}^{\mu \nu} + r' B_{\mu \nu} \widetilde{B}'^{\mu \nu} )+ \sin \theta_\chi \sum_a W^a_{\mu \nu} \widetilde{W}^{a \mu \nu} +... \right),
\ee
where the ellipsis denote additional couplings to $B' B'$ which are unimportant for our purposes.\footnote{Also, for Majorana DM, CP-even couplings to $\overline{\chi} \chi$ would lead to $p$-wave suppressed annihilations.}  As in \Refs{Weiner:2012cb,Weiner:2012gm}, the overall scale of the operator is
\be
\frac{1}{\Lambda^3_R} = \frac{g^2 \lambda^2}{48 M_f^3 \pi^2},
\ee
where $\lambda$ is the Yukawa coupling between DM and the messengers, and we have made the simplifying assumption that the scalar and fermion messenger masses are the same.  

Considering the usual annihilation scenario with a DM mass $M_\chi = 130~\GeV$, we find the annihilation cross section is approximately
\begin{eqnarray}
\vev{\sigma v}_s (\chi \chi \rightarrow \gamma \gamma) & = & \frac{\alpha^2_\lambda \alpha^2_W (\cos \theta_\chi \cos^2 \theta_W+\sin \theta_\chi \sin^2 \theta_W)^2}{9 \pi M_\chi^2} \left(\frac{M_\chi}{M_M}\right)^6\\
& \simeq &  1.3 \times 10^{-27} \, \text{cm}^3  / \text{s} \left( \frac{\alpha_\lambda}{1.8} \right)^{2} \left(\frac{ 200 \text{ GeV}}{M_M}\right)^6 \left(\frac{M_\chi}{ 130 \text{ GeV}} \right)^4. \nonumber
\end{eqnarray}
where we have chosen messengers with the electroweak quantum numbers of the Higgs, such that $\cos \theta_\chi = 0.29$.  Hence quite a strongly-coupled, but still perturbative, theory is required unless the charged messenger mass is reduced close to the DM mass.  

Alternatively, with the addition of a $Z'$, the \Fermi line at $130~\GeV$ could result from DM semi-annihilation with a larger DM mass.  In this case, the semi-annihilation cross section is
\begin{eqnarray}
\frac{1}{2} \vev{\sigma v}_s (\chi \chi \rightarrow \gamma Z') &  = &  \frac{\alpha^2_\lambda \alpha^2_W (\cos \theta_\chi \cos \theta_W r')^2}{36 \pi M_\chi^2} \left(\frac{M_\chi}{M_M}\right)^6 \left( 1- \frac{M^2_{Z'}}{4 M_\chi^2} \right)^3\\
 & \simeq &   2.3 \times 10^{-27} \, \text{cm}^3  / \text{s} \left(\frac{r'}{10} \right)^2 \left(\frac{\alpha_\lambda}{0.72} \right)^{2} \left(\frac{ 200 \text{ GeV}}{M_M}\right)^6   \nonumber \\
& &  \times \left(\frac{M_\chi}{ 173 \text{ GeV}} \right) \left(\frac{E_\gamma}{ 130 \text{ GeV}} \right)^3 \nonumber
\end{eqnarray}
where $M_{Z'} = 2 M_\chi \sqrt{1-E_\gamma/M_\chi}$ has been replaced such that $E_\gamma = 130$ GeV is ensured for any choice of $M_\chi$.\footnote{This explains the linear scaling with $M_\chi$, since the kinematic factor $(1-M_{Z'}^ 2/ 4 M_\chi^2)^3 = (E_\gamma / M_{\chi})^3$.}  Again, this benchmark cross section is a factor of $(173/130)^2$ larger than the nominal $1.3\times10^{-27} \, \text{cm}^3  / \text{s}$ value required for annihilating DM \cite{Weniger:2012tx}.  While a value $r' = 10$ might seem on the large side, this is only due to the small size of the electroweak couplings, and $r' = 10$ simply corresponds to a $\U(1)'$ coupling $\alpha' \sim 1$.

By using this retrofitting technique, semi-annihilation can extend more restricted annihilation models to allow for a wider range of DM masses and smaller couplings.  It should be kept in mind that a line at $112~\GeV$ from the $\gamma Z$ final state is no longer a prediction of the model, though it can be easily accommodated if necessary by having multiple $Z'$ states.  Also, we have not addressed the question of relic abundance, however the observed abundance can again be easily explained by extending the dark sector to include e.g.\  a dark singlet scalar mixed with the SM Higgs.

\section{Conclusions}
\label{sec:conclude}

Regardless of whether the tentative $130~\GeV$ \Fermi gamma line persists, this intriguing feature is a reminder that monochromatic galactic photons offer a unique probe of DM, and it is therefore important to understand DM scenarios in which such gamma lines can be enhanced.  In this paper, we have shown that semi-annihilation $\psi_i \psi_j \to \psi_k \gamma$ can yield parametrically larger cross sections for gamma lines than annihilation $\psi_i \overline{\psi}_i \to \gamma \gamma$.  This makes semi-annihilation a well-motivated target for future gamma line studies.  Furthermore, unlike explanations of the the \Fermi gamma line based on DM annihilation, semi-annihilation allows for a wide range of DM masses, presenting the opportunity for DM to impact indirect detection experiments at a variety of scales.

We have highlighted two interesting semi-annihilation scenarios.  First, we considered a vector DM scenario with a custodial symmetry which generates a $130~\GeV$ line from semi-annihilation of $173~\GeV$ DM.  Finding a sub-dominant $173~\GeV$ gamma line from annihilation to $\gamma \gamma$ would be a smoking gun for this scenario, as would proving the \emph{absence} of a $\gamma Z$ line at $112~\GeV$.  In the context of an $\SU(3)_d$ dark gauge group, various group-theoretic factors conspire to require a larger than expected gauge coupling to explain the \Fermi line, but there are indeed benchmark values of the messenger mass and dark gauge coupling where one still has perturbative control.

Second, we have shown how existing annihilation models can be retrofitted into semi-annihilation models.  Focusing on the case of RayDM, explaining the \Fermi line with the annihilation diagram $\chi \chi \to \gamma \gamma$ via the Rayleigh operator requires large (but still perturbative) Yukawa couplings to messengers.  We have shown that including an additional massive dark $Z'$ allows the semi-annihilation process $\chi \chi \to Z' \gamma$ to achieve the same $130~\GeV$ feature for smaller couplings and a wider range of DM masses.  

As we eagerly await confirmation of the $130~\GeV$ line, semi-annihilation is a reminder that multiple gamma lines offer the ability to perform DM spectroscopy.  If the dark sector is sufficiently rich, then $N$ species of DM can yield as many as $\mathcal{O}(N^3)$ lines through a combination of annihilation and semi-annihilation, with strengths determined by the various dark couplings and symmetries.  Such gamma line measurements would then offer crucial insights into the nature of the dark sector.

\acknowledgments{We thank Raffaele Tito D'Agnolo, Douglas Finkbeiner, Mariangela Lisanti, Yasunori Nomura, Maurizio Pierini, and Jacob Wacker for helpful conversations. This work is supported by the U.S. Department of Energy (DOE) under cooperative research agreement DE-FG02-05ER-41360.  F.D. is supported by the Miller Institute for Basic Research in Science, M.M. is supported by a Simons Postdoctoral Fellowship, and J.T. is supported by the DOE Early Career research program DE-FG02-11ER-41741.}

\bibliographystyle{JHEP}
\bibliography{SEMIFERMI}

\end{document}